# Vacancy enhanced cation ordering enables >15% efficiency in Kesterite solar cells

Jinlin Wang[1,2]†, Licheng Lou[1,2]†, Kang Yin[1,2]†, Fanqi Meng[3]†, Xiao Xu[1], Menghan Jiao[1,2], Bowen Zhang[1,2], Jiangjian Shi[1]*, Huijue Wu[1], Yanhong Luo[1,2,5], Dongmei Li[1,2,5]*, Qingbo Meng[1,4,5]*

[1]Beijing National Laboratory for Condensed Matter Physics, Renewable Energy Laboratory, Institute of Physics, Chinese Academy of Sciences (CAS); Beijing, 100190, P. R. China.

[2]School of Physical Sciences, University of Chinese Academy of Sciences; Beijing, 100049, P. R. China.

[3]School of Materials Science and Engineering, Peking University; Beijing, 100871, P. R. China.

[4]Center of Materials Science and Optoelectronics Engineering, University of Chinese Academy of Sciences; Beijing, 100049, P. R. China.

[5]Songshan Lake Materials Laboratory; Dongguan, 523808, P. R. China.

*Corresponding author. Email: shijj@iphy.ac.cn; dmli@iphy.ac.cn; qbmeng@iphy.ac.cn.

†These authors contributed equally to this work.

**Abstract**

Atomic disorder, a widespread problem in compound crystalline materials, is a imperative affecting the performance of multi-chalcogenide $Cu_2ZnSn(S, Se)_4$ (CZTSSe) photovoltaic device known for its low cost and environmental friendliness. Cu-Zn disorder is particularly abundantly present in CZTSSe due to its extraordinarily low formation energy, having induced high-concentration deep defects and severe charge loss, while its regulation remains challenging due to the contradiction between disorder-order phase transition thermodynamics and atom-interchange kinetics. Herein, through introducing more vacancies in the CZTSSe surface, we explored a vacancy-assisted strategy to reduce the atom-interchange barrier limit to facilitate the Cu-Zn ordering kinetic process. The improvement in the Cu-Zn order degree has significantly reduced the charge loss in the device and helped us realize 15.4% (certified at 14.9%) and 13.5% efficiency (certified at 13.3%) in 0.27 $cm^2$ and 1.1 $cm^2$-area CZTSSe solar cells, respectively, thus bringing substantial advancement for emerging inorganic thin-film photovoltaics.

Entropy-driven atomic disordering is a crucial concern that impacts the precise fabrication and functional design of crystalline materials[1,2]. This issue is particularly prominent in the photovoltaic material known for its advantages such as low cost and environmental friendliness, the multinary chalcogenide $Cu_2ZnSn(S, Se)_4$ (CZTSSe), because it owns the most complex element components studied so far[3-5]. CZTSSe comprises three metal cations of similar size and identical coordination structure in its lattice (i.e. Cu, Zn and Sn). The disorder or reciprocal substitution between these cations triggers intense phase competition, band fluctuations, and deep defects, thereby impinging on the photoelectric conversion performance of CZTSSe solar cells[6-9]. Among them, the Cu-Zn disorder is abundantly present due to its extraordinarily low thermodynamic formation energy[10,11], resulting in the reduction of the material's effective band gap[6,12,13], and an increase in the concentration and charge capturing rate of deep defects[6,14,15].

A variety of efforts have been undertaken to address the Cu-Zn disorder issue in CZTSSe. Specifically, researchers developed a range of cation substitution methods to increase the thermodynamic formation energy of Cu-Zn disorder[16-20]. Yet, whether the occupation of external cation can improve the disorder of intrinsic atoms in the material remains contentious[21,22]. Researchers also started from the order-disorder phase-transition thermodynamics of the binary system, exploring a long-time low-temperature post-annealing strategy[8,13,23]. While an increased degree of ordering atomic and band-edge electronic structure were observed, this approach has not yet obtained effective device performance improvement[21,24,25], probably because the long-time annealing also induced other negative effects such as the surface atom volatilization loss and surface reconstruction. Overall, these years of continued efforts have yielded little substantial success. This outcome has led to a stalemate in the investigation and regulation of Cu-Zn disorder, and even induce doubts of whether Cu-Zn disorder significantly impairs the material quality and device performance. Nevertheless, despite the challenges in experiment, recent theoretical studies have once again demonstrated a strong correlation between Cu-Zn disorder and the charge loss in CZTSSe[26], thereby reaffirming the necessity and urgency of addressing the Cu-Zn disorder issue.

Herein, we focus more on the surface of CZTSSe film since it is believed to have a more significant impact on the solar cell performance. Starting from the perspective of atomic ordering kinetics, we have developed a vacancy-assisted method to reduce the atom migration barrier for facilitating the Cu-Zn atomic ordering process. The vacancy on the CZTSSe film surface was artificially introduced via ion pre-doping and subsequent solution etching. The success of improving the surface Cu-Zn ordering has effectively mitigated interface defects and carrier non-radiative recombination, and also promoted charge transport. Ultimately, we achieved a record efficiency of 15.4% (certified at 14.9%) in Kesterite solar cells, as well as a certified efficiency of >13% on a 1.1 $cm^2$-area cell. This result represents significant advancement for emerging inorganic thin-film solar cells and moreover furnishes opportunities for more precise disorder-order control in crystalline materials.

**Experimentally introducing surface atomic vacancies**

It was predicted that under thermodynamic equilibrium, lower temperature would help the Cu-Zn binary system transform to a higher degree of order. However, from the perspective of kinetics, the system at lower temperatures does not have sufficient energy to drive the direct interchange of Cu-Zn atoms (schematically shown in Figure 1A). This contradiction between thermodynamics and kinetics can be an important reason that limits the improvement of Cu-Zn ordering and cell performance.

Generally, the atom ordering kinetic rate is proportional to $\exp(-E_B/K_BT)$ ($E_B$ is the energy barrier of atom interchange, $K_B$ is the Boltzmann constant and $T$ is the temperature)[27-29]. This suggests that reducing the $E_B$ can promote the atomic ordering process kinetically while without sacrificing the final atomic ordering degree of the system. Vacancy mechanism is a promising way to realize this purpose as vacancies provide larger space to accommodate the interchanged atoms, as shown in Figure 1A. Whereas the direct interchange of atoms is primarily realized through interstitial path, which is usually accompanied by large lattice deformation. We conducted first-principles calculations to estimate the $E_B$ for Cu-Zn ordering processes via direct or vacancy-assisted interchange mechanism. For the direct interchange, the $E_A$ reached 2.8 eV (Figure 1B). While through the vacancy mechanism (Cu vacancy), $E_A$ was significantly reduced to 1.3 eV, indicating that the kinetic rate of the ordering process can be enhanced by more than 16 orders of magnitude. This means that the vacancy mechanism can facilitate the enhancement of Cu-Zn ordering both thermodynamically and kinetically.

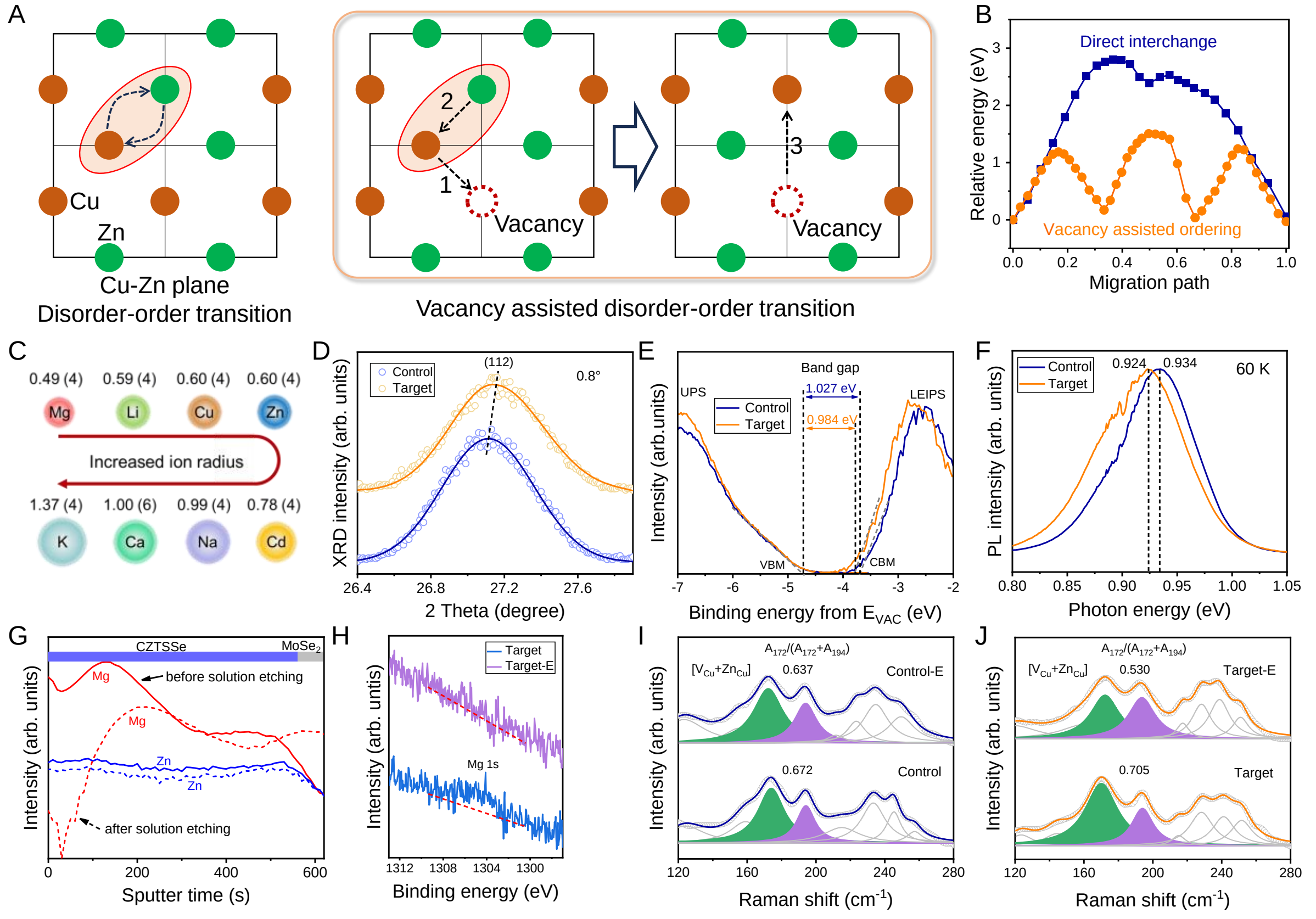

**Figure1. Artificial introduction of atomic vacancies.** (A) Schematic diagram of Cu-Zn interchange processes, via direct or Cu vacancy ($V_{Cu}$) assisted mechanism. (B) Corresponding energy profiles along the interchange paths. (C) Ionic radii[30] of several elements that can be used to form cation doping in Kesterite lattice. (D) Grazing incidence X-Ray diffraction (112) peak, (E) UPS valence and LEIPS conduction band-edge electronic structure and (F) steady-state PL spectra of the pristine CZTSSe films. (G) Elemental profiles of Zn and Mg and (H) Mg 1s XPS of the Mg-doped CZTSSe films before and after solution etching. Raman spectra (excited at 325 nm) of control (I) and Mg-doped (J) CZTSSe films before and after solution etching.

The key to realize the vacancy-assisted atom ordering in experiment is to introduce sufficient benign vacancies in the region of interest in the CZTSSe film. We explored an ion pre-doping and subsequent solution etching strategy to artificially introduce vacancies in the film surface region. Group IA and IIA elements are potential candidates for the doping in CZTSSe (Figure 1C). In our experimental approach, the chlorides of these elements were firstly deposited onto the surface of CZTSSe precursor films. After the selenization, the films were then etched by ammonia solutions to dissolve the doped ions. The films were subsequently annealed for about one hour in a vacuum chamber (Ar, 0.2 Pa) to drive the Cu-Zn ordering process. Solar cells based on these films were finally fabricated after CdS, ZnO, ITO (Sn doped $In_2O_3$) and Ni/Al depositions. Judging from the final solar cell performance (Figure S1-6), Li, Na and Cd can only improve the average PCE by only ~0.1%, while K and Ca were detrimental to the PCE. Mg element was found to be the champion choice, which improved the average photoelectric conversion efficiency (PCE) of the cell from ~13% to ~14.8%. Moreover, thermal evaporation deposition of $MgF_2$ on the precursor film surface and spin-coating with a $MgCl_2$ solution yielded similar outcomes (Figure S7). These results can be explained as the ion size of Na, K and Ca is obviously larger and thus cannot effectively dope at Cu or Zn sites[31-33]. For Li and Cd, it was found that they can hardly be dissolved from the film (Figure S8). In addition, Cd mainly dopes at the Zn site[34,35] while the ordering through the Zn vacancy has much higher $E_B$ (~1.9 eV) than that through the Cu vacancy (Figure S9). For Mg, previous researches demonstrated that it could dope at both Cu ($Mg_{Cu}$) or Zn ($Mg_{Zn}$) site and in some cases the $Mg_{Cu}$ even exhibited negative formation energy[33,36,37]. From the perspective of Coulombic attraction and charge compensation, $Mg_{Cu}$ could also prefer to form close to the $Cu_{Zn}$ site within the Cu-Zn disorder pair. These properties of Mg ion and its induced $V_{Cu}$ are very conducive to the vacancy-assisted ordering process depicted in Figure 1 A. For clarity, in the following, the undoped or the Mg-doped sample will be denoted as control or target sample, respectively; the samples after being solution etched will be denoted as control (or target)-E; the etched and annealed samples will be denoted as control (or target)-E-A.

We further characterized the Mg doping and its induced $V_{Cu}$ in the film surface. Grazing incidence X-Ray diffraction found that the Mg incorporation made the (112) peak of CZTSSe slightly shift to a higher angle (Figure 1D). Surface band-edge electronic structure measurement using ultraviolet photoelectron spectroscopy (UPS) and low-energy inverse photoelectron spectroscopy (LEIPS) demonstrated that the Mg incorporation slightly reduced the bandgap of CZTSSe (Figure 1E). Correspondingly, the steady-state photoluminescence (PL) of the doped film also showed a little bathochromic shift (Figure 1F). Further combined with elemental characterization (Figure 1G-H), we confirmed that Mg element had indeed doped into the CZTSSe lattice.

The elemental characterization shown in Figure 1G-H further found that Mg was significantly removed from the film surface through the solution etching. The ICP-OES (inductively coupled plasma optical emission spectrometer) analysis of the etching solution also demonstrated that Mg was dissolved (Figure S10). We further used the Raman peak intensity ratio ($R$=$A_{172}$/($A_{172}$+$A_{194}$), excited at 325 nm) to investigate the $V_{Cu}$ properties of these films. In this analysis, lower $R$ value stands for higher $V_{Cu}$ concentration[38,39]. As in Figure 1I, the $R$ value of the control sample was only a little reduced by the solution etching, indicating that the $V_{Cu}$ concentration was slightly increased. Comparatively, a significant decrease of the $R$ was observed in the target sample, indicating an obvious increase in the $V_{Cu}$ concentration. In addition, the $R$ value of the etched target sample was also obviously smaller than that of the etched control sample, implying that the Mg pre-doping and solution etching had effectively introduced $V_{Cu}$ in the film surface.

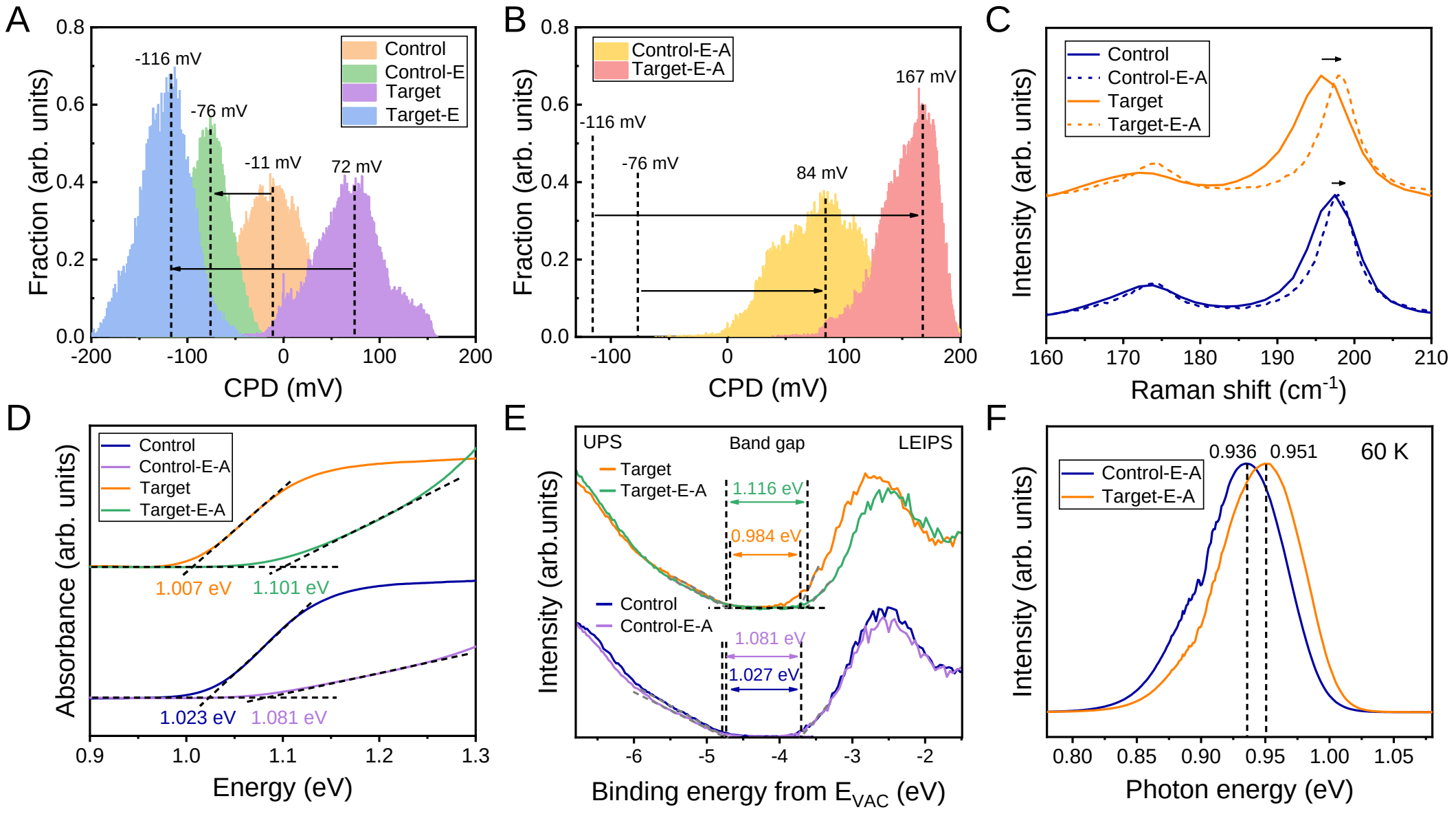

**Figure 2. Vacancy enhanced atomic ordering**. (A) CPD of the control and target films before and after solution etching. (B) CPD of the control and target films after being annealed. Raman spectra (C), UV-Visible absorption spectra transformed from reflection mode (D), UPS and LEIPS spectra (E) of the films before and after experiencing the atomic ordering process. (F) Steady-state PL spectra of the control-E-A and target-E-A films.

## Vacancy enhanced Cu-Zn atomic ordering

We further demonstrated the annealing induced Cu-Zn ordering in the CZTSSe lattice. Using

Kelvin-probe force microscopy, the surface contacting potential difference (CPD) of these films were firstly measured (Figure 2A-B and S11). As in Figure 2A, the solution etching process moved CPD of both the control and target films to lower values, and especially, the CPD shift of the target film reached ~190 mV, further supporting the effective formation of $V_{Cu}$ acceptors. After the annealing, CPD of both the films moved to higher values (Figure 2B), indicating a reduction in the density of acceptors such as $V_{Cu}$ and $Cu_{Zn}$. Cu-Zn ordering process could be an important cause to reduce $Cu_{Zn}$ acceptors[23].

More characteristic evidence for the improvement in the Cu-Zn ordering in the target-E-A film came from the Raman spectra[8,40,41]. As in Figure 2C, compared to the as-selenized target film, the solution etching and annealing processes obviously shifted the Raman peak of Kesterite phase to a higher wavenumber position, also accompanied by a narrowing of the Raman peak. Comparatively, without Mg pre-doping, the solution etching and annealing processes had little influence on the Raman spectra of the control sample. Optical bandgap of CZTSSe is another signature related to the Cu-Zn ordering degree[8,12,13], which was evaluated by reflectance spectroscopy in our investigations (Figure 2D and S12). The result exhibited that after the etching and annealing process, the bandgap of the target sample increased more significantly than that of the control sample, reached ~95 meV (Figure 2D). This change in the bandgap was further demonstrated by UPS and LEIPS. As seen in Figure 2E, the bandgap of target film surface increased by ~130 meV, realized by the shift of both the valence and the conduction band edge. This was also reflected in the steady-state PL spectra. The PL peak of the target-E-A sample shifted to 0.951 eV (Figure 2F), significantly larger than that of the control-E-A and pristine target samples (Figure 1F). Overall, these characterization results demonstrated that our proposed strategy (Mg pre-doping, solution etching and annealing) had effectively improved the Cu-Zn ordering of CZTSSe films.

**Defect and charge loss characterization**

We further investigated the influence of improved Cu-Zn ordering on the defect and charge loss properties of the fabricated Kesterite solar cells. The energetic distribution and charge capturing properties of defects in the cells were firstly characterized using thermally admittance spectroscopy (TAS) (Figure 3A-B and S13). As the Arrhenius plots shown in Figure 3A, although the ionization energy ($E_t$) of defects in the two samples did not exhibit essential difference (possibly $Cu_{Zn}$ defect in both samples)[10,42,43], the defect attempt-to-escape frequency ($v_0$) of the target sample was found to be ~15 times smaller than that of the control sample, indicating that the defect charge capturing velocity was significantly reduced. This should be a result of the increased atomic ordering, which reduced the electron-lattice vibration interaction in the solid material.

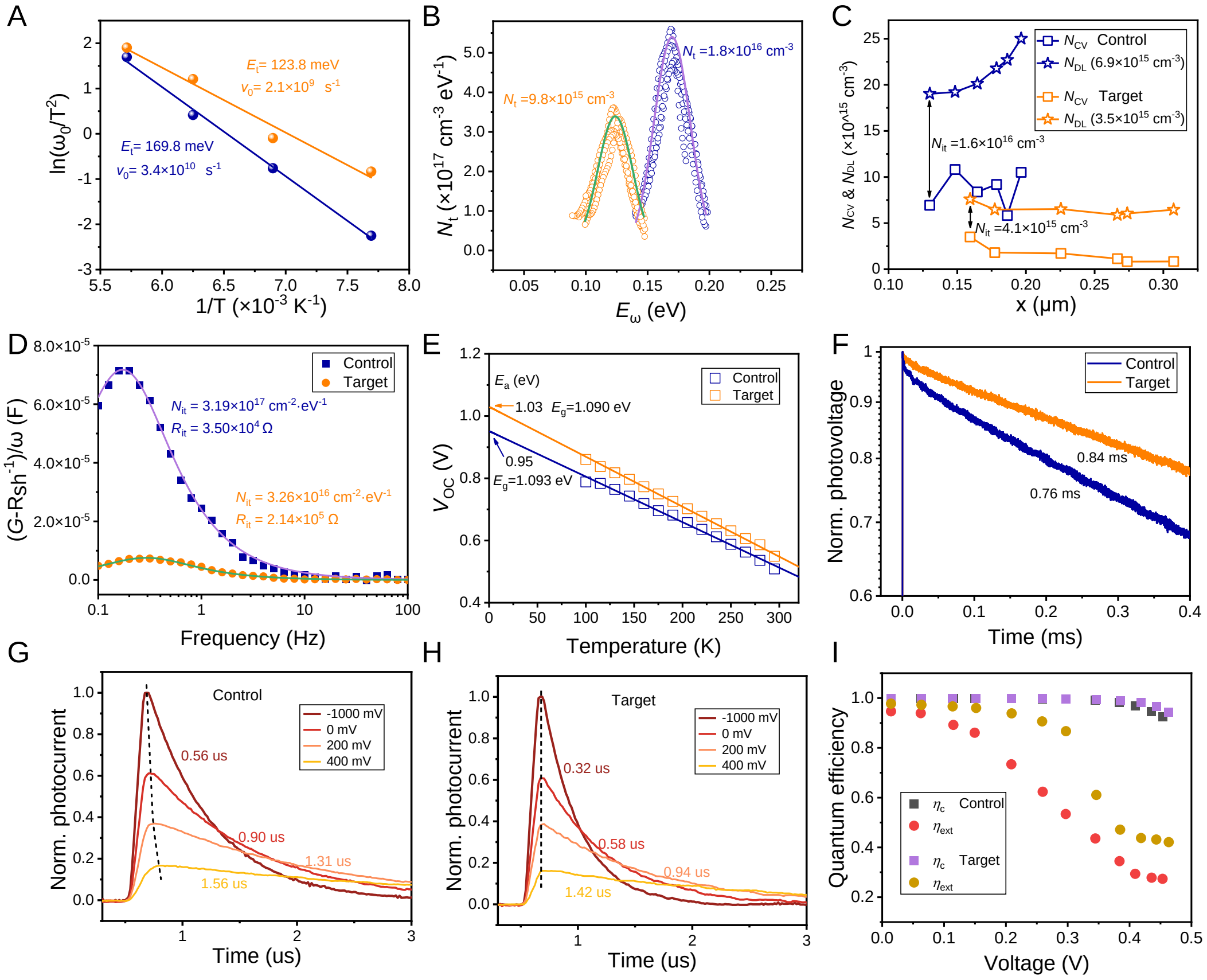

**Figure 3. Defect and charge loss characterization of the cells.** (A) Arrhenius plots of charge capturing and (B) the energetic distribution of defects (Gaussian-type) in the cells. (C) Charge profiles in the CZTSSe absorber layer derived from capacitance-voltage ($N_{CV}$) and drive-level charge profiling ($N_{DL}$) measurements. (D) Quantification of the interface defect in the cell using frequency-dependent admittance ($G$) analysis. $R_{sh}$: shunt resistance of the cell, $\omega$: Angular frequency. (E) Temperature-dependent open-circuit voltage ($V_{OC}$) of the cells. (F) Photovoltage decay dynamics of the cells at 0 V. (G-H) Photocurrent decay dynamics of the cells at different bias voltages. The dashed lines depicted the evolution of photocurrent peaks. (I) Charge extraction ($\eta_{ext}$) and collection ($\eta_C$) efficiencies of the cells derived from the modulated electrical transient measurements.

We further quantified the defect density using different capacitance analysis methods. Firstly, TAS results exhibited that the total density of $Cu_{Zn}$ defect in the target sample was only 1/2 of that in the control sample, directly demonstrating the reduced Cu-Zn disordering (Figure 3B). More specific spatial distribution of the defect charge was measured by capacitance-voltage (CV) and drive-level capacitance profiling (DLCP) methods. As in Figure 3C, the target sample exhibited much lower charge concentration in the measured region due to the reduced $Cu_{Zn}$ acceptors. Particularly, the interface defect density of the target sample ($N_{it}$, estimated from the difference in the charge density measured by CV and DLCP) was only one forth of that in the control sample. We also used an extended physics model to investigate the charge capturing properties of interface defects (Figure S14)[44]. Through fitting the frequency-dependent admittance curve, the density of interface defects that participated in the charge capturing ($N_{it}$) in the target sample was estimated to be $3.26\times10^{16}$ $cm^{-2}eV^{-1}$, which was only 1/10 of that in the control sample. The fitting result also indicated that the target sample had much higher resistance of interface charge capturing ($R_{it}$), that is, more difficult interface charge capturing.

The aforementioned defect characterization demonstrated that the density and charge capturing velocity of defects, especially the interface defect, have been significantly suppressed through the Cu-Zn ordering process. As a result, the target cell obtained more superior open-circuit voltage ($V_{OC}$) at temperatures ranging from 300 to 100 K and moreover higher recombination activation energy, $E_A$, than that of the control cell (Figure 3E and S15). The very small discrepancy between $E_A$ and optical bandgap indicated that the fast sub-gap relaxation of photo-generated carriers had become negligible in the target CZTSSe sample, also proving that localized electronic states caused by the Cu-Zn disordering had been significantly suppressed[6,7,21].

We also characterized the charge dynamics of the cells using modulated electrical transient methods[45,46]. As in Figure 3F, the target cell exhibited much longer photovoltage decay lifetime and moreover obviously reduced fast decay in the early stage. This phenomenon implies that the interface charge recombination in the cell has been suppressed, agreeing well with the defect characterization. For the transient photocurrent, the target cell exhibited obviously smaller decay time than the control cell. In addition, the photocurrent peak position of the target cell almost kept constant at different bias voltages, significantly distinct to the control cell whose photocurrent peak moved continuously with the voltage increase. These phenomena indicated that the target cell had better charge transport ability. The elimination of localized sub-gap electronic states and their induced charge trapping and detrapping processes is an important beneficial reason for this result. The improved charge transport ability should be able to help more photoinduced carriers in the CZTSSe absorber be extracted into the buffer/window layer, especially when the built-in electric field was reduced at high voltages. This was confirmed by the higher charge extraction efficiency ($\eta_{ext}$) quantified from the modulated electrical transient results (Figure 3I)[47]. In addition, due to the reduce interface charge recombination, the extracted charge in the window layer can also be collected by the electrodes more efficiently, resulting in higher charge collection efficiency ($\eta_C$). The reduction in bulk and interfacial charge loss is expected to improve PCE of the cell.

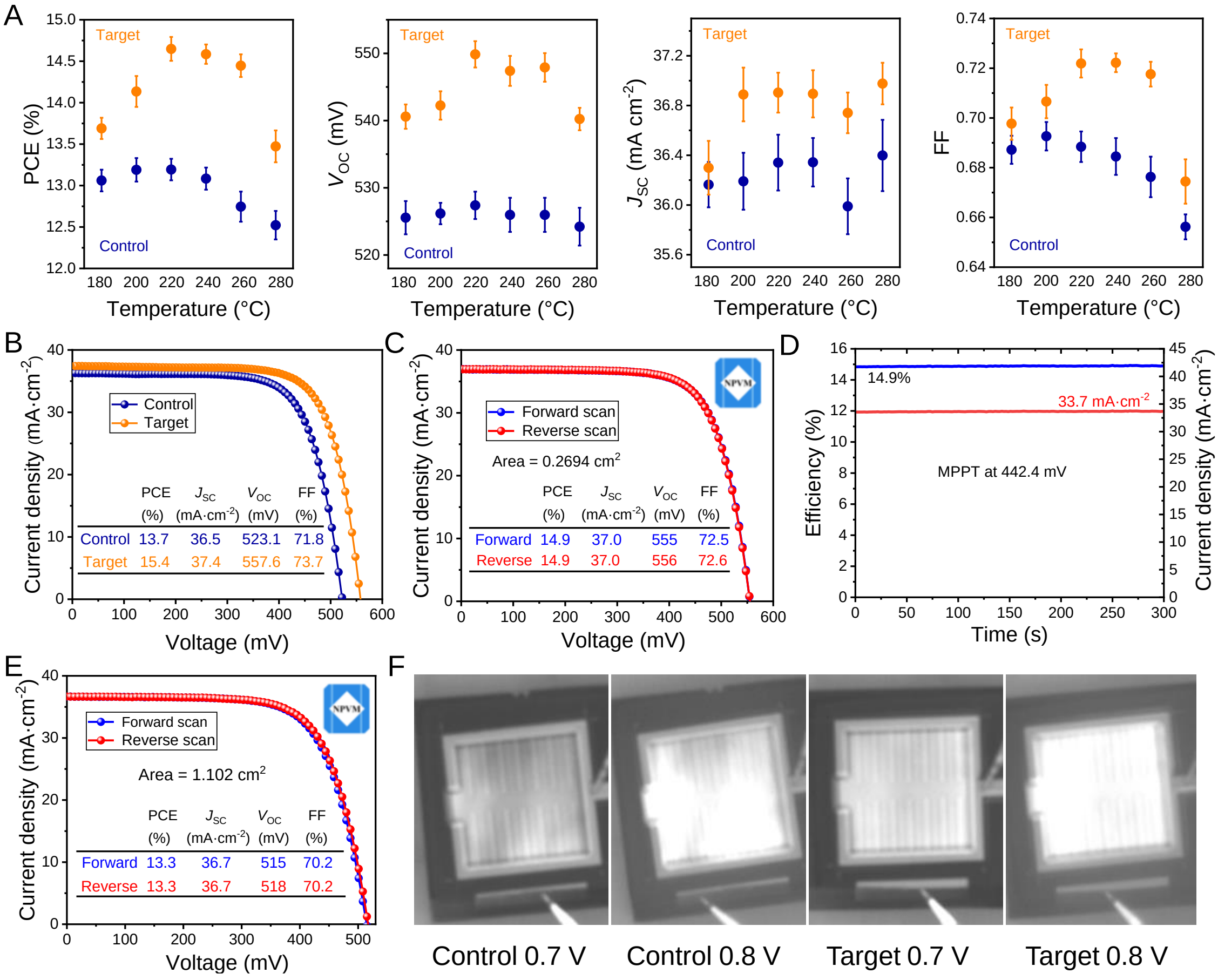

**Figure 4. Solar cell performance.** (A) Statistical performance parameters of the cells with CZTSSe films annealed at different nominal temperatures. Error bar: standard deviation. (B)

Current density-voltage (*J*-*V*) characteristics of champion devices. (C) Certified *J*-*V* and (D) maximum power point tracking curves (300 s) of the champion target device (0.27 $cm^2$). (E) Certified *J*-*V* curves of the large-area device (1.1 $cm^2$). (F) Electroluminescence images of the control and target devices.

**Device performance characterization**

To obtain the optimal Cu-Zn ordering result, we have systematically optimized the Mg pre-doping concentration and the subsequent annealing temperature. The performance parameters of the cells obtained from different nominal annealing temperatures are shown in Figure 4A. As can be seen, average PCE of the target cell was enhanced from ~13.7% to ~14.6% when the temperature increased from 180 to 220 °C. This enhancement was reflected in all three performance parameters, $V_{OC}$, short circuit current ($J_{SC}$) and fill factor (FF). Particularly, the average FF was improved from ~0.69 to ~0.72, and the average $V_{OC}$ reached 550 mV in the optimal group, confirming the reduced charge recombination in the cell. Comparatively, the PCE of the control cell was only enhanced by ~0.1% (from 13.1% to 13.2%). This result agrees well with the fact that the atomic ordering was limited in the control cell.

The current density-voltage (*J*-*V*) characteristics of the champion cells are given in Figure 4B. The target cell achieved a high PCE of 15.4% with $J_{SC}$ of 37.4 mA $cm^{-2}$, $V_{OC}$ of 557.6 mV and FF of 0.737. All these performance parameters are much superior to that of the control cell which exhibited a PCE of 13.7%. One of the cells in the target group was also sent to an accredited independent laboratory (National PV Industry Measurement and Testing Center, NPVM) for certification. The certified PCE (*J*-*V* measurement, Figure 4C) reached 14.9% (Figure S16), representing a substantial advancement for emerging inorganic thin-film solar cells. The cell also exhibited excellent operational stability in the maximum power point tracking (MPPT) process, also demonstrating a steady-state PCE of 14.9% (Figure 4D).

We also fabricated 1.1 $cm^2$-area solar cells and their certified PCE reached 13.3% (both *J*-*V* and MPPT, Figure 4E and Figure S17-18), also a new record in this field. The realization of this performance primarily benefited from high uniformity of photoelectric properties of the cell at the centimeter scale, as demonstrated by the electroluminescence (EL) characterization [48,49]. As shown in Figure 4F, the target cell exhibited uniform EL in the whole active region, while only part region of the control cell can be light up under charge injection. Therefore, our introduced vacancy strategy not only enables more efficient atomic ordering, but also enables this process to occur more synchronously at different spatial locations. These benefits allow this strategy to continue to play constructive roles in preparing larger-sized CZTSSe solar cells, such as Kesterite photovoltaic modules.

**Acknowledgements**

This work was supported by the National Natural Science Foundation of China (Grant nos. U2002216 (Q. M.), 52222212 (J. S.), 52172261 (Y. L.), 52227803 (Q. M.), 51972332 (H.W.)). J. S. also gratefully acknowledges the support from the Youth Innovation Promotion Association of the Chinese Academy of Sciences (2022006).

**Author contributions**

Jinlin Wang, Jiangjian Shi, Dongmei Li and Qingbo Meng conceived the idea and designed the experiments. Jinlin Wang and Jiangjian Shi did the experiments and the data analysis. Xiao Xu, Menghan Jiao and Bowen Zhang supported CZTSSe solar cell fabrication. Fanqi Meng performed STEM measurements. Licheng Lou and Kang Yin performed DLCP and PL measurements. Huijue Wu and Yanhong Luo supported M-TPC/TPV characterization and discussions. Jinlin Wang, Jiangjian Shi, Dongmei Li and Qingbo Meng participated in writing the manuscript.